\documentstyle[aps,preprint]{revtex}
\begin{document}
\draft
 
\title{Multipole Expansion for Relativistic Coulomb Excitation}
\author{H. Esbensen}
\address{Physics Division, Argonne National Laboratory, Argonne, 
Illinois 60439}
\author{C. A. Bertulani}
\address{Physics Department, Brookhaven National Laboratory,  
Upton, NY 11973-5000 - USA}
\date{\today}
\maketitle

\begin{abstract}
We derive a general expression for the multipole expansion of the 
electro-magnetic interaction in relativistic heavy-ion collisions, 
which can be employed in higher-order dynamical calculations of Coulomb
excitation.  The interaction has diagonal as well as off-diagonal 
multipole components, associated with the intrinsic and relative
coordinates of projectile and target.  
A simple truncation in the off-diagonal components gives excellent 
results in first-order perturbation theory for distant collisions
and for beam energies up to 200 MeV/nucleon.
\end{abstract}
 

\narrowtext

\section{Introduction}

The Coulomb excitation of nuclei at relativistic energies is commonly
calculated in first-order perturbation theory, using the formalism
developed in Ref. \cite{AW}. The derivation is restricted to distant
collision, where the intrinsic radial coordinates are smaller than the 
minimum distance between the colliding nuclei.
In calculations of the Coulomb dissociation of proton halo nuclei,
it is of interest also to consider close collisions, since the 
density of the valence proton can extend to very large distances. 
Moreover, higher-order processes may also play a role, as suggested 
by the non-relativistic calculations of the $^8$B breakup reported 
in Ref. \cite{EB96}.

In order to calculate the Coulomb dissociation of halo nuclei at 
energies where relativistic and higher-order effects cannot be 
ignored, one would need realistic multipole form factors for the
electro-magnetic interaction. Such form factors have recently been 
studied \cite{cab99} but they were only determined in the 
long-wave-length limit for distant collisions. We therefore find it 
timely to study the multipole decomposition of the interaction at 
relativistic energies, both for close and distant collisions.
The form factors we derive can be applied in higher-order dynamical 
calculations of two-body breakup reactions, such as the continuum
discretized coupled-channels calculations \cite{Nunes}, or in
numerical methods that evolve the two-body wave function in coordinate
space \cite{EBB,Kido,Mele}.

We base our study on the so-called Li\'enard-Wiechert potential
and include the effect of the convection current. The magnetization 
current, on the other hand, is ignored for simplicity. 
We assume that the relative velocity of the interacting nuclei is a
constant, so that the retardation effect associated with a
time-dependent velocity can be ignored.

The multipole expansion of the interaction contains both diagonal and 
off-diagonal multipole components. The off-diagonal components appear
because of the Lorentz contraction. Our approach is similar to the 
method developed by Baltz et al. \cite{Baltz} in their study of the 
$e^+e^-$ pair production in relativistic heavy-ion collisions.   
We extend their study and derive the multipole fields that are 
relevant to Coulomb excitation.

The multipole expansion of the Li\'enard-Wiechert potential
is presented in Sect. II. There we also discuss how to calculate the
interaction associated with the convection current.
We propose in Sect. III a simple truncation in the the sum over 
off-diagonal multipole components. We test this approximation in
Sect. IV in first-order perturbation theory for distant collisions 
by comparing to the exact results of Ref. \cite{AW}. 
Sect. V contains our conclusions.

\section{Electro-magnetic interaction with fast charged particle}

The electro-magnetic interaction of a fast charged particle (with atomic
number $Z_1$) and a weakly bound proton in a target nucleus has the
form \cite{AW}
$$V_{\rm em}({\bf r},{\bf r}') = 
Z_1e^2\Bigl(\phi-{{\bf v}\over 2mc^2}
[\hat{\bf p}\phi+\phi\hat{\bf p}]\Bigr).\eqno(1)$$
The charged particle is assumed to move with constant velocity {\bf v}
in the z-direction on the straight-line trajectory 
${\bf r}'={\bf b}+{\bf v}t$, with impact parameter {\bf b} with respect
to the target nucleus.
The coordinates of the proton in the target nucleus are denoted by
{\bf r}, and $\hat{\bf p}$ is the conjugate momentum operator.
The potential $\phi$ is the Li\'enard-Wiechert potential
$$\phi={\gamma\over \sqrt{|{\bf r}_\perp-{\bf r}'_\perp|^2
+\gamma^2(z-z')^2}},\eqno(2)$$
where $\gamma=1/\sqrt{1-\beta^2}$, and $\beta=v/c$. 
The second part of the interaction (1), which contains the momentum
operator $\hat{\bf p}$, is due to the convection current. We ignore
the magnetization current but it could be included by replacing the 
momentum operator by $\hat{\bf p} + \hbar
\mbox{\boldmath$\nabla$} \times \mbox{\boldmath$\mu$}_p$, 
where $\mbox{\boldmath$\mu$}_p$ is the magnetic moment of the proton.

In the two-body breakup of a target nucleus, into a proton and a core 
nucleus, one would actually have to consider the effect of two 
interactions with the projectile, namely, the `direct' interaction 
with the proton and the `recoil' interaction with the core.   
We will not discuss this complication here, since the multipole 
expansion of the two interactions can easily be generated from the 
formulation given below.

\subsection{Multipole expansion}

The multipole expansion of the Li\'enard-Wiechert potential $\phi$ can 
be obtained from the Fourier representation \cite{Baltz}
$$\phi = {4\pi\over (2\pi)^3} \int d{\bf q} \ 
{e^{i{\bf q}\cdot ({\bf r}-{\bf r}')} \over q_\perp^2+q_z^2/\gamma^2}
= {4\pi\over (2\pi)^3} \int d{\bf q} \ 
{e^{i{\bf q}\cdot ({\bf r}-{\bf r}')} \over q^2}
\ {1\over 1-\beta^2 \ \cos^2(\theta_q)}.\eqno(3)$$
Except for the last factor, this is just the ordinary Coulomb potential. 
To proceed, we introduce the multipole expansion of the last 
term in Eq. (3) 
$${1\over 1 - \beta^2 \cos^2(\theta)} = 
\sum_{\lambda \ {\rm even}} g_\lambda(\beta)
\ P_\lambda(\cos(\theta_q)),\eqno(4a)$$
where 
$$g_\lambda(\beta) =
(2\lambda+1) \ \beta^{-1} \ Q_\lambda(\beta^{-1}),\eqno(4b)$$
and 
$$Q_\lambda(z)={1\over 2}\int_{-1}^1 dt \ 
{P_\lambda(t)\over z-t}\eqno(4c)$$ 
are Legendre polynomials of the second kind; 
cf. Eq. 8.825 of Ref. \cite{GR} or Eq. (8.8.3) of Ref. \cite{Abra}.
Explicit expressions and recursion relations for $Q_\lambda(z)$ are 
given in Ref. \cite{Abra}.

We can now insert the plane wave expansion
$$e^{i{\bf q\cdot r}} = 4\pi \sum_{lm} i^l \ j_l(qr) \
Y_{lm}({\hat q}) Y_{lm}^*({\hat r})$$
for the two plane waves in Eq. (3) and obtain
$$\phi = 4\pi \sum_{lml'm'} i^{l-l'} \
R_{ll'}(r,r') \ A_{lm,l'm'}(\beta) \
Y_{lm}^*({\hat r}) \ Y_{l'm'}({\hat r}'),\eqno(5)$$
where
$$R_{l,l'}(r,r') = {2\over \pi}  
\int_0^\infty dq \ j_l(qr) \ j_{l'}(qr')=
{1\over \sqrt{rr'}} \
\int_0^\infty {dq\over q}
J_{l+1/2}(qr) \ J_{l'+1/2}(qr'),\eqno(5a)$$ 
and
$$A_{lm,l'm'}(\beta) =
\langle Y_{l'm'}| {1\over 1-\beta^2\cos^2(\theta_q)} | Y_{lm}\rangle
= \sum_{\lambda \ {\rm even}} g_\lambda(\beta) \
\langle Y_{l'm'}({\hat q}) | P_\lambda(\theta_q) |
Y_{lm}({\hat q})\rangle.\eqno(5b)$$

The above expressions were derived for a trajectory with constant
velocity in the z-direction.  The matrix (5b) is therefore diagonal 
in {\it m}. Let us also give a more general result which does not 
refer to any specific coordinate system. This can be done by 
noticing that the angle $\theta_q$ in Eq. (5b) is actually the angle 
between {\bf q} and the velocity {\bf v}.  Thus by inserting 
$P_\lambda(\theta_q)$ = 
$\sum_\mu D_{\mu 0}^\lambda({\hat v})^* D_{\mu 0}^\lambda({\hat q})$
we obtain
$$A_{lm,l'm'}(\beta,{\hat v}) =
\sum_{\lambda \ {\rm ever}} g_\lambda(\beta) \
\sum_\mu \Bigl(D_{\mu 0}^\lambda({\hat v})\Bigr)^* \
\langle lm\lambda\mu|l'm'\rangle \
\langle l'0\lambda 0|l0\rangle,\eqno(5c)$$
where the last two Clebsch-Gordan coefficients is the matrix element
$\langle Y_{l'm'}| D_{\mu 0}^\lambda| Y_{lm}\rangle$.

We note that the expansion (5) contains diagonal (i.e. $l=l'$) as well
as off-diagonal multipole components (i.~e. $(lm)\neq(l'm')$. 
The off-diagonal components are caused by the Lorentz contraction which 
destroys the spherical symmetry of the potential $\phi$. 
The radial dependence is determined by the integral (5a), and analytic
expressions and recursion relations are derived in App. A.
In numerical applications one can include the finite size of the 
projectile simply by multiplying the integrand in Eq. (5a) by the 
Fourier transform of the charge distribution. However, we find it 
instructive to show the analytic expressions for
point-particles. 

The radial dependence of the diagonal components is determined by
(see Eq. (A.5))
$$R_{l,l}(r,r') ={1\over 2l+1} \ {r_<^l\over r_>^{l+1}},\eqno(6)$$
where $r_<=\min(r,r')$ and $r_>=\max(r,r')$. This is identical to the 
dependence one has in the non-relativistic limit. In fact, the 
well-known multipole expansion of the Coulomb interaction is recovered 
from Eq. (5), when we insert the non-relativistic limit of Eq. (5b), 
viz.  $A_{lm,l'm'}(\beta=0)$ = $\delta_{ll'}\delta_{mm'}$.

The off-diagonal multipole contributions to $\phi$ are more complicated.
The radial form factors have the symmetry property
$$R_{l,l'}(r,r') = R_{l',l}(r',r).\eqno(7a)$$
According to Eq. (5b), we only need expressions for even values of 
$|l-l'|$. We show in App. A that
$$R_{l,l+\Lambda}(r,r') = 
0, \ \ \ {\rm for} \ \ r\geq r' \ \ \ {\rm and} \ \ 
\Lambda=2,4,...\eqno(7b)$$
The explicit, non-zero expressions can be obtained from Eq. (A.4).
For $\Lambda=2$ we obtain in particular (see Eq. (A.5))
$$R_{l,l+2}(r,r') = {1\over 2} \ {r^l\over r'^{l+1}} \
[1-({r\over r'})^2], \ \ {\rm for} \ \ r<r'.\eqno(8)$$

One can exploit the properties (7a) and (7b) and write expressions for 
$\phi$ that are valid for distant and close collisions, respectively.
Thus we obtain for distant collisions ($r<r'$)
$$\phi_{\rm dist} = \sum_{lmm'} 4\pi Y_{lm}^*({\hat r})
\sum_{\Lambda=0,2,.}
i^\Lambda \ R_{l,l+\Lambda}(r,r') \ A_{lm,l+\Lambda m'} \ 
Y_{l+\Lambda,m'}({\hat r}').\eqno(9a)$$
For close collisions ($r>r'$) we can use the expression
$$\phi_{\rm close} = \sum_{lmm'} 4\pi Y_{lm}^*({\hat r})
\sum_{\Lambda=0,2,.}
i^\Lambda \ R_{l,l-\Lambda}(r,r') \ A_{lm,l-\Lambda m'} \ 
Y_{l-\Lambda,m'}({\hat r}').\eqno(9b)$$
The sum over $\Lambda$ is finite for close collisions, since 
$l-\Lambda$ must be non-negative. This is a very nice feature, 
which makes it feasible to include all terms.
For distant collisions, on the other hand, one would have to 
make a truncation in the sum over $\Lambda$.  

\subsection{Convection current interaction}

The contribution to the interaction (1) that originates from the
convection current is
$$\phi' = {i\hbar\over 2mc^2} \Bigl(
({\bf v}\cdot\mbox{\boldmath$\nabla$})\phi \ + \
\phi ({\bf v}\cdot\mbox{\boldmath$\nabla$})\Bigr)
= {i\hbar\over 2mc^2} \Bigl(
{\bf v}\cdot(\mbox{\boldmath$\nabla$}\phi) \ + \
2 \phi \ ({\bf v}\cdot\mbox{\boldmath$\nabla$})\Bigr).\eqno(10)$$
The operator ${\bf v}\cdot\mbox{\boldmath$\nabla$}$ is effectively of 
dipole nature, 
as shown in App. B. In practical numerical calculations it will 
operate on an expression of the form $r^{-1} f(r) Y_{LM}({\hat r})$. 
>From Eq. (B.4) we then obtain
$${\bf v}\cdot\mbox{\boldmath$\nabla$} \Bigl({f(r)\over r} 
Y_{LM}({\hat r})\Bigr) = 
v \sum_\mu \Bigl(D_{\mu 0}^1({\hat v})\Bigr)^*
\sum_{L'M'} Y_{L'M'}({\hat r}) \
\langle Y_{L'M'}|D_{\mu 0}^1 | Y_{LM} \rangle$$
$$\times
\Bigl({1\over r}{df(r)\over dr} - {f(r)\over 2r^2} \
[L'(L'+1)-L(L+1)]\Bigr).\eqno(11)$$

We can also give an explicit expression for the term 
$c^{-1}{\bf v}\cdot(\mbox{\boldmath$\nabla$} \phi)$, which appears 
in second version of Eq. (10).
Assuming that {\bf v} points in the z-direction, we obtain
from Eq. (3)
$$\beta({d\phi\over dz}) =
{4\pi\over (2\pi)^3} \int d{\bf q} \
{e^{i{\bf q}({\bf r}-{\bf r}')}\over q^2} \
{i\beta q\cos(\theta_q)\over 1-\beta^2\cos^2(\theta_q)}.$$
We can now repeat the derivation that lead to Eq. (5) and obtain
$$\beta\left({\partial \phi \over \partial z}\right)= 
4\pi i \sum_{ll'} i^{l-l'} \
P_{ll'}(r,r') \sum_m B_{ll'm}(\beta) \
Y_{lm}^*({\hat r}) \ Y_{l'm}({\hat r}'),\eqno(12)$$
where
$$P_{ll'}(r,r') = {2\over \pi}  
\int_0^\infty dq \ q \ j_l(qr) \ j_{l'}(qr'),\eqno(12a)$$ 
and
$$B_{ll'm}(\beta) =
\langle Y_{l'm}| {\beta\cos(\theta)\over 1-\beta^2\cos^2(\theta)} | 
Y_{lm}\rangle
= \sum_{\lambda \ {\rm odd}} g_\lambda(\beta)\
\langle Y_{l'm} | P_\lambda(\cos(\theta)) | Y_{lm} \rangle.
\eqno(12b).$$
The last expression is similar to Eq. (5b) but the sum 
is now over odd values of $\lambda$.
The dependence on the radial coordinates can be derived from Eqs.
(A.7-8). We note that matrix elements of $\phi'$ can be calculated
in a much simpler way in first-order perturbation theory,
as we shall see in Sect. IV.

\section{Significance of relativistic corrections}

In order to illustrate the significance of relativistic corrections
at intermediate energies, we show in Fig. 1 the functions 
$g_\lambda(\beta)$ defined in Eq. (4b). They are seen to decrease
rapidly as function of $\lambda$ when $(v/c)^2<0.5$. This suggests
that one would only need a few terms in the sum over $\Lambda$
in Eq. (9a).

In the next section we determine the range of velocities where a
truncation to $\Lambda=0,2$ is reasonable for distant collision. 
Thus we will apply the approximation 
$$\phi_{lm}^{\rm dist} \approx {4\pi\over 2l+1} \ {r^l\over r'^{l+1}} \ 
Y_{lm}^*({\hat r})$$
$$\times\sum_{m'} \Bigl[A_{lm,lm'}(\beta) \ 
Y_{lm'}({\hat r}') -
{2l+1\over 2} \ A_{lm,l+2,m'}(\beta) \
Y_{l+2,m'}({\hat r}') \ (1-(r/r')^2)\Bigr].\eqno(13)$$
We assume a straight line trajectory with constant velocity in
the z-direction and calculate the Coulomb excitation in first-order
perturbation theory. We can then test the accuracy of the truncation,
Eq. (13), by comparing to the exact results from Ref. \cite{AW}.

\section{Test in first-order perturbation theory}

In first-order perturbation theory for distant collisions one needs the 
Fourier integral of the multipole fields,
$$\phi_{lm}({\bf r},\omega) = \int_{-\infty}^\infty dt \ e^{i\omega t} 
\phi_{lm}({\bf r},t)
= \sqrt{4\pi\over 2l+1} \ r^l \ Y_{lm}^*({\hat r}) \ 
S_{lm}(\omega),\eqno(14a)$$
where $\hbar\omega$ is the excitation energy and $S_{lm}(\omega)$ are 
the so-called orbital integrals,
$$S_{lm}(\omega) = \sqrt{4\pi\over 2l+1} \ 
\int_{-\infty}^\infty dt \ e^{i\omega t} \ 
{Y_{lm}({\hat R})\over R^{l+1}}.\eqno(14b)$$
For non-relativistic Coulomb excitation and a straight-line
trajectory, these integrals are \cite{AW}
$$S_{lm}^{\rm NR}(\omega) = 
{2\over v} \ {i^{l+m}\over \ 
\sqrt{(l+m)! (l-m)!}} \ ({\omega\over v})^l \
K_m(\omega b/v).\eqno(15)$$
Below we derive the orbital integrals for the truncated interaction
(13).

\subsection{Li\'enard-Wiechert potential}

The Fourier transform (14a) of the multipole field (13) can be
expressed in terms of various combinations of the non-relativistic 
orbital integrals (15). This is obvious for the first term in (13),
and the second term is calculated in App. C.
A slight complication is the very last term in Eq. (13) which
contains the factor $(r/r')^2$. It turns out to be a factor
of $(\omega r/v)^2$ smaller than the dominant term so let us
ignore it. Thus we obtain
$$S_{lm}(\omega) = A_{llm}(\beta) \ S_{lm}^{\rm NR}(\omega)
- A_{l,l+2,m}(\beta) \ {2l+1\over 2} 
% \Bigl[SS_{l2m} - 
% r^2 \sqrt{2l+5\over 2l+1} \
% S_{l+2,m}^{\rm NR}(\omega)\Bigr],$$
% where
% $$SS_{l2m}(\omega) = 
\sqrt{4\pi\over 2l+1} \ 
\int_{-\infty}^\infty dt \ e^{i\omega t} \ 
{Y_{l+2,m}({\hat R})\over R^{l+1}}.\eqno(16)$$
The latter integral is given in Eq. (C.5).

\subsection{Convection current interaction}

We also need to calculate the contribution from the convection 
current, Eq. (10). Here we can again make the substitution (B.1).
Assuming that the single-particle potential commutes with ${\bf r}$
we can go a step further and replace the momentum operator by
$p_z = {im\over \hbar} \ [H,z]$, where $H$ is the single-particle
Hamiltonian. Thus we obtain
$$\phi' = -{i\beta\over 2\hbar c}
\Bigl([H,z]\phi + \phi [H,z]\Bigr) 
= -{i\beta\over 2\hbar c}
\Bigl([H,z\phi] + \phi Hz -zH\phi\Bigr).\eqno(17)$$ 
We note that in first-order perturbation theory one can effectively 
replace the commutator $[H,z\phi]$ by $\hbar\omega z\phi$.
The other term, $\phi Hz - zH\phi$, has to be considered more
carefully.

Let us write the multipole expansion of $\phi'$ as in Eq. (14a),
$$\phi' = \sqrt{4\pi\over 2l+1} r^l \ Y_{lm}({\hat r}) \ S'_{lm}.$$
Since $z$ in Eq. (17) is a dipole operator, we see that $\phi'_{lm}$ 
can receive contributions from both $\phi_{l-1,m}$ and $\phi_{l+1,m}$. 
The contribution from the latter is smaller (by a factor of 
$(\omega r/v)^2$) so let us ignore it. Let us first consider
the contribution from the commutator $[H,z\phi]$. This leads to
the expression
$$S'_{lm}([H,z\phi]) \approx 
-{i\beta^2\over 2} \ {\omega\over v} \ S_{l-1,m}
\ {\sqrt{l^2-m^2}\over 2l-1}.\eqno(18)$$

For the dipole field we see that the second term in Eq. (17), namely 
$\phi_{00} Hz-zH\phi_{00}$, is identical to the first term because 
$\phi_{00}$ commutes with the Hamiltonian {\it H}. Thus we obtain
$$S'_{10} \approx -\beta^2 \ {i\omega\over c} S_{00}, \ \ \ 
S'_{11}=0.\eqno(18a)$$
For the quadrupole field we can neglect the second term of Eq. (17)
It is identical to zero when we insert $\phi_{10}$, whereas 
inserting $\phi_{11}$ leads to a magnetic type transition
($\propto (l_x+il_y)$). Thus we will use the approximation
$$S'_{2m} \approx
-{i\beta^2\over 2} \ {\omega\over v} \ S_{1,m}
\ {\sqrt{4-m^2}\over 3}.\eqno(18b)$$

\subsection{Comparison to exact results} 

In order to illustrate the orbital integrals we obtain from the 
approximation (13), it is useful to plot the reduced values 
$${\widetilde S}_{lm} = {S_{lm}+S'_{lm}\over N_{lm}}, \ \ \
{\rm where} \ \
N_{lm} = {2\over v} \ {i^{l+m} \ b^{-l}\over
\sqrt{(l+m)!(l-m)!}}.\eqno(19)$$
The results we obtain for dipole and quadrupole excitations are shown 
by the solid symbols in Figs. 2 and 3, respectively, as functions of 
the adiabaticity parameter $\xi=\omega b/v$.
Here we have chosen the beam velocity $v/c=0.6$ 

The full relativistic expression for electric transitions \cite{AW},
which also includes a contribution from the convection current, is 
$${\widetilde S}_{lm}^{\rm Rel}(\omega) = 
S_{lm}^{\rm Rel}/N_{lm} =
{1\over \gamma} \ G_{lm}(\beta) \ 
\xi^l \ K_m({\xi\over\gamma}),\eqno(20)$$
where $G_{lm}$ can be extracted from \cite{AW},
$$G_{10} = G_{20} = G_{2\pm 2} = {1\over\gamma}, \
G_{1\pm 1} = 1, \ G_{2\pm 1} = 1-{\beta^2\over 2}.$$
This result is shown by the solid curves in Figs. 2 and 3. 
The dashed curves represent the non-relativistic result
${\widetilde S}_{lm}^{NR} = \ \xi^l \ K_{m}(\xi)$, c.~f. Eq. (15).
It is seen that relativistic effects are significant and that the 
truncated calculation is in very good agreement with the full result.
If we choose a larger velocity, say $v/c$ = 0.8, some discrepancy
starts to occur, so it will be necessary to include higher values of
$\Lambda$ in the sum (9a). 

\section{Conclusion}

The multipole expansion of the electro-magnetic interaction in 
relativistic heavy-ion collisions that we have presented is exact 
for a straight-line trajectory with constant velocity. It has not 
previously, as far as we know, been used in calculations of the
Coulomb excitation. The formulation is, in particular, applicable 
to reactions where close collisions play a role, and it is also
well suited for calculations of higher-order processes in the
Coulomb dissociation of halo nuclei at intermediate energies. 

The multipole expansion contains diagonal as well as off-diagonal
components, associated with the intrinsic coordinates and the
coordinates for the relative motion of the colliding nuclei.
The off-diagonal multipole components are caused by relativistic
effects, essentially by the Lorentz contraction of the Coulomb field;
only the diagonal components survive in the non-relativistic limit.

The number of off-diagonal multipole components is finite for close 
collisions but it is, in principle, infinite for distant collisions.
However, it is sufficient to include just a few off-diagonal terms
at intermediate energies.
Thus we find that the diagonal and first off-diagonal terms 
(i.~e., $\Lambda=0,2$ in Eq. (9a)) give excellent results in 
first-order perturbation theory for distant collisions and for
velocities up to $v/c\approx 0.6$, i.~e. up to about 200 MeV/nucleon.

{\bf Acknowledgments} This work was supported by the US Department
of Energy, Nuclear Physics Division, under Contracts W-31-109-ENG-38
and DE-AC02-98CH10886, 
and by a fellowship (C.A.B.) from the 
J.S. Guggenheim Memorial Foundation. 

\section{Appendices}

\subsection{Bessel function integrals}

Here we study the integral in Eq. (5a)
$$I_{\mu,\mu+2m}(a,b) = \int_0^\infty {dq\over q} \
J_\mu(aq) \ J_{\mu+2m}(bq),\eqno(A.1)$$
where $\mu=l+1/2$ and $2m$ is an even number, according to 
parity selection of Eq. (5b).
This integral is given in many textbooks but we shall mainly 
refer to Abramowitz \cite{Abra}. 
We note that it has the symmetry
$$I_{\mu+2m,\mu}(a,b) = I_{\mu,\mu+2m}(b,a),\eqno(A.2)$$
so it is sufficient to determine (A.1) for all values of $a$ and $b$.

Let us first assume that $a<b$. From Eq. 11.4.34 of Ref. \cite{Abra}
we obtain (with $\nu=\mu+2m$ and $\lambda=1$)
$$I_{\mu,\mu+2m}(a,b) = {1\over 2} ({a\over b})^\mu \
{\Gamma(\mu+m)\over \Gamma(\mu+1) \ \Gamma(m+1)}
\ _2F_1(\mu+m,\ -m;\ \mu+1; \ (a/b)^2).\eqno(A.3)$$
The hypergeometric function in Eq. (A.3) is a polynomial of degree 
{\it m}, which can be obtained directly for the series Eq. 15.1.1 
of Ref. \cite{Abra}. Inserting into that expression the Pochhammer's 
symbols 
% $(a)_n = \Gamma(a+n)/\Gamma(a)$ and 
% $(-m)_n=(-1)^n \Gamma(m+1)/\Gamma(m+1-n)$
(see Eq. (6.1.22) of Ref. \cite{Abra}) we obtain
$$I_{\mu,\mu+2m}(a,b) = {1\over 2} ({a\over b})^\mu \sum_{n=0}^m
{(-1)^n \ \Gamma(\mu+m+n)\over \Gamma(m+1-n)\Gamma(\mu+1+n)} \
{(a/b)^{2n}\over n!}.\eqno(A.4)$$
For $m$ = 0, 1 and 2 we obtain in particular
$$I_{\mu,\mu}(a,b) = {1\over 2\mu}(a/b)^\mu, \ \ \ \
I_{\mu,\mu+2}(a,b) = {1\over 2} (a/b)^\mu
\Bigl[1-(a/b)^2\Bigr],$$
$$I_{\mu,\mu+4}(a,b) = {1\over 4} (a/b)^\mu \ \Bigl[1-(a/b)^2\Bigr] \
\Bigl[\mu+1-(\mu+3)(a/b)^2\Bigr].\eqno(A.5)$$
Using the symmetry property (A.2) we see that 
$I_{\mu,\mu}$ is consistent with Eq. (6). 

When $a>b$ and $m$ is non-zero we find that
$$I_{\mu,\mu+2m}(a,b) = 0, \ \ \ {\rm for} \ \ a>b,
\ \ \ m=1,2,...\eqno(A.6)$$
This result follows from Eq. 11.4.33 of Ref. \cite{Abra} because
that expression will contain the factor
$1/\Gamma(-m+1)$ which is zero for $m=1,2,3,...$

The functions $P_{l,l'}(r,r')$ defined in Eq. (12a) can be obtained
from the relation
$$P_{l,l'+1}(r,r') + P_{l,l'-1}(r,r') =
{2l'+1\over r'} \ R_{l,l'}(r,r').\eqno(A.7)$$
This expression follows directly from the definitions in Eqs. (5a) and 
(12a) and the recursion relation for spherical Bessel functions,
$zj_{l-1}(z)+zj_{l+1}(z)$ = $(2l+1)j_l(z)$.
For $l'=l+1$ we find explicitly (c.f. Eq. 6.575 of Ref. \cite{GR})
$$ P_{l,l+1}(r,r') = P_{l+1,l}(r',r) = 
{1\over r'^2} ({r\over r'})^l, \ \ \ {\rm for} \ r\leq r',\eqno(A.8)$$
while it is zero when $r>r'$. The radial dependence can now in general
be derived from the recursion relation (A.7).

\subsection{Current operator}

Here we derive an expression for the operator
${\bf v}\cdot\mbox{\boldmath$\nabla$}$,
which appears in the contribution from the convection current.
It can be expressed in terms of a commutator with the Laplace operator
$\Delta$ as follows
$${\bf v}\cdot\mbox{\boldmath$\nabla$} = {1\over 2}
[\Delta,{\bf v}\cdot{\bf r}].\eqno(B.1)$$
Here we can insert
$${\bf v}\cdot{\bf r} = v r \cos(\theta_{vr}) =
v r \sum_\mu \Bigl(D_{\mu 0}^1({\hat v})\Bigr)^*
\ D_{\mu 0}^1({\hat r}),\eqno(B.2)$$
and express the Laplace operator in spherical coordinates
$$\Delta = {d^2\over dr^2} +{2\over r}{d\over dr} -
{{\hat L}^2\over r^2},\eqno(B.3)$$
where ${\hat L}$ is the angular momentum operator.
Thus we obtain
$${\bf v}\cdot\mbox{\boldmath$\nabla$} = v
\sum_\mu \Bigl(D_{\mu 0}^\lambda({\hat v})\Bigr)^*
\Bigl( D_{\mu 0}^\lambda({\hat r})({d\over dr} + {1\over r})
-{1\over 2r} 
[{\hat L}^2, D_{\mu 0}^\lambda({\hat r})]\Bigr).\eqno(B.4)$$

\subsection{Orbital integral}

Here we derive an expression for the integral
$${\cal S}_{l2m}(\omega) = 
\sqrt{4\pi\over 2l+1} \ 
\int_{-\infty}^\infty dt \ e^{i\omega t} \ 
{Y_{l+2,m}({\hat R})\over R^{l+1}},\eqno(C.1)$$
which appears in Eq (16). This integral can be calculated in a 
simple way by observing that
$$Y_{l+2,m} = \sqrt{2l+5\over 2l+3} \
\sum_{\mu=-1}^1 C_\mu^1 \ D_{\mu 0}^1({\hat R}) \
Y_{l+1,m-\mu}({\hat R}),\eqno(C.2)$$
where (use Eq. (14) of App. D in Ref. \cite{AWB})
$$C_0^1 = {\sqrt{(l+2+m)(l+2-m)}\over l+2},$$
$$C_{-1}^1 = {1\over \sqrt{2}} \ {\sqrt{(l+1-m)(l+2-m)}\over l+2},$$
$$C_1^1 = {1\over \sqrt{2}} \ {\sqrt{(l+1+m)(l+2+m)}\over l+2}.$$

The trajectory we consider falls in the x-z plane, with
$\cos(\theta) = vt/R$ and $\sin(\theta) = b/R$.
The integral (C.1) can therefore be expressed in terms of the 
non-relativistic orbital integrals (15) as follows
$${\cal S}_{l2m}(\omega) = \sqrt{2l+5\over 2l+1} \
\Bigl[ C_0^1 \ v(-i{d\over d\omega}) \ S_{l+1,m}^{NR}(\omega)$$
$$+ {b\over \sqrt{2}} \Bigl( C_{-1}^1 S_{l+1,m+1}^{NR}(\omega) -
C_1^1 \ S_{l+1,m-1}^{NR}(\omega)\Bigr)\Bigr].\eqno(C.3)$$

Working out the details this can be written explicitly as
$${\cal S}_{l2m}(\omega) = {1\over v} \
\sqrt{2l+5\over 2l+1} \ {1\over l+2} \
{i^{l+2+m}\over \sqrt{(l+2+m)!(l+2-m)!}} \ ({\omega\over v})^l$$
$$\Bigl[(2l+3)\xi\Bigl((l+2-m)K_{m+1}(\xi) + (l+2+m)K_{m-1}(\xi)\Bigr)
-2(l+1)(l+2+m)(l+2-m) K_m(\xi)\Bigr],\eqno(C.4)$$
where $\xi=\omega b/v$. Or even more compact:
$${\cal S}_{l2m}(\omega) = -{2\over v} \
\sqrt{2l+5\over 2l+1} \ 
{i^{l+2+m}\over \sqrt{(l+2+m)!(l+2-m)!}} \ ({\omega\over v})^l$$
$$\Bigl[\Bigl((l+1)(l+2)+m^2\Bigr)K_m(\xi)
-{2l+3\over 2} \ \xi \Bigl(K_{m-1}(\xi)+K_{m+1}(\xi)\Bigr)\Bigr].
\eqno(C.5)$$

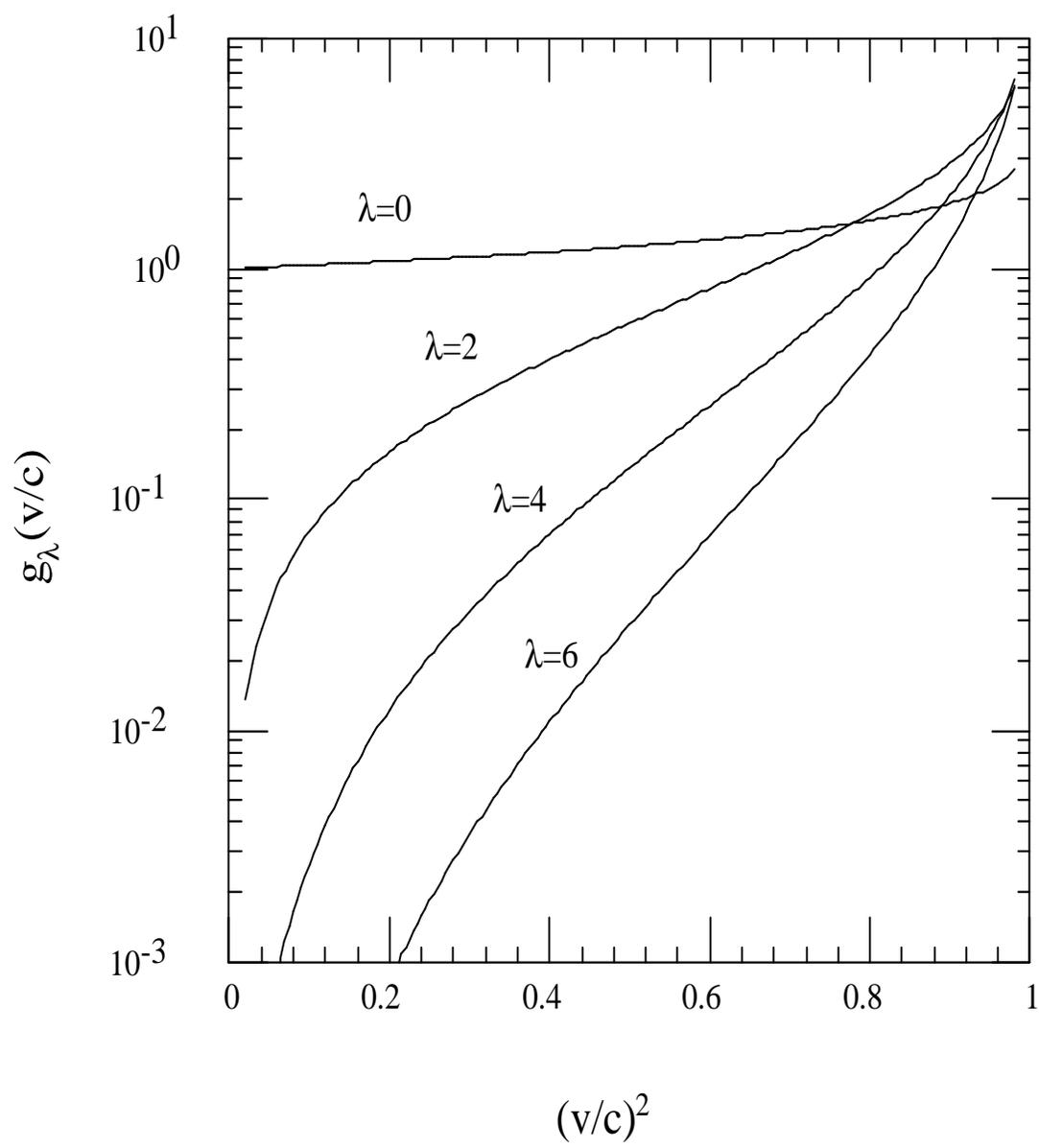
\begin{figure}
\caption{The functions $g_\lambda(v/c)$, defined in Eq. (4b), are shown
as functions of $(v/c)^2$ for $\lambda$ = 0-6.}
\label{autonum1}
\end{figure} 

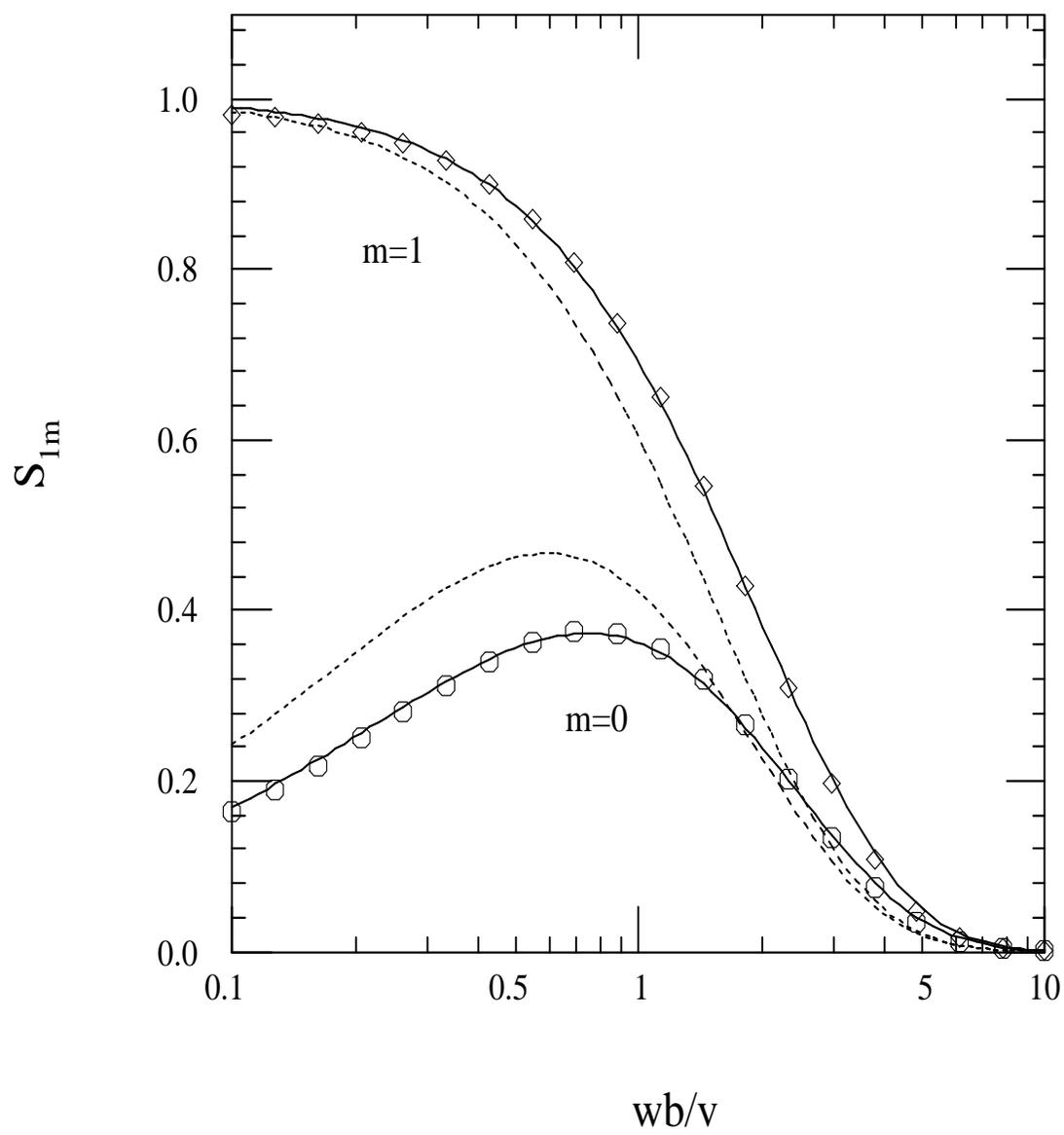
\begin{figure}
\caption{The reduced values of orbital integrals, defined in Eq. 
(19), are shown for dipole excitations as functions of 
the adiabaticity parameter $\xi=\omega b/v$.
The beam velocity is $v/c=0.6$. The solid curves are the exact results,
Eq. (20), and the dashed curves are the non-relativistic results.
The solid symbols are the results of our approximation discussed
in Sect. IV.}
\label{autonum2}
\end{figure} 

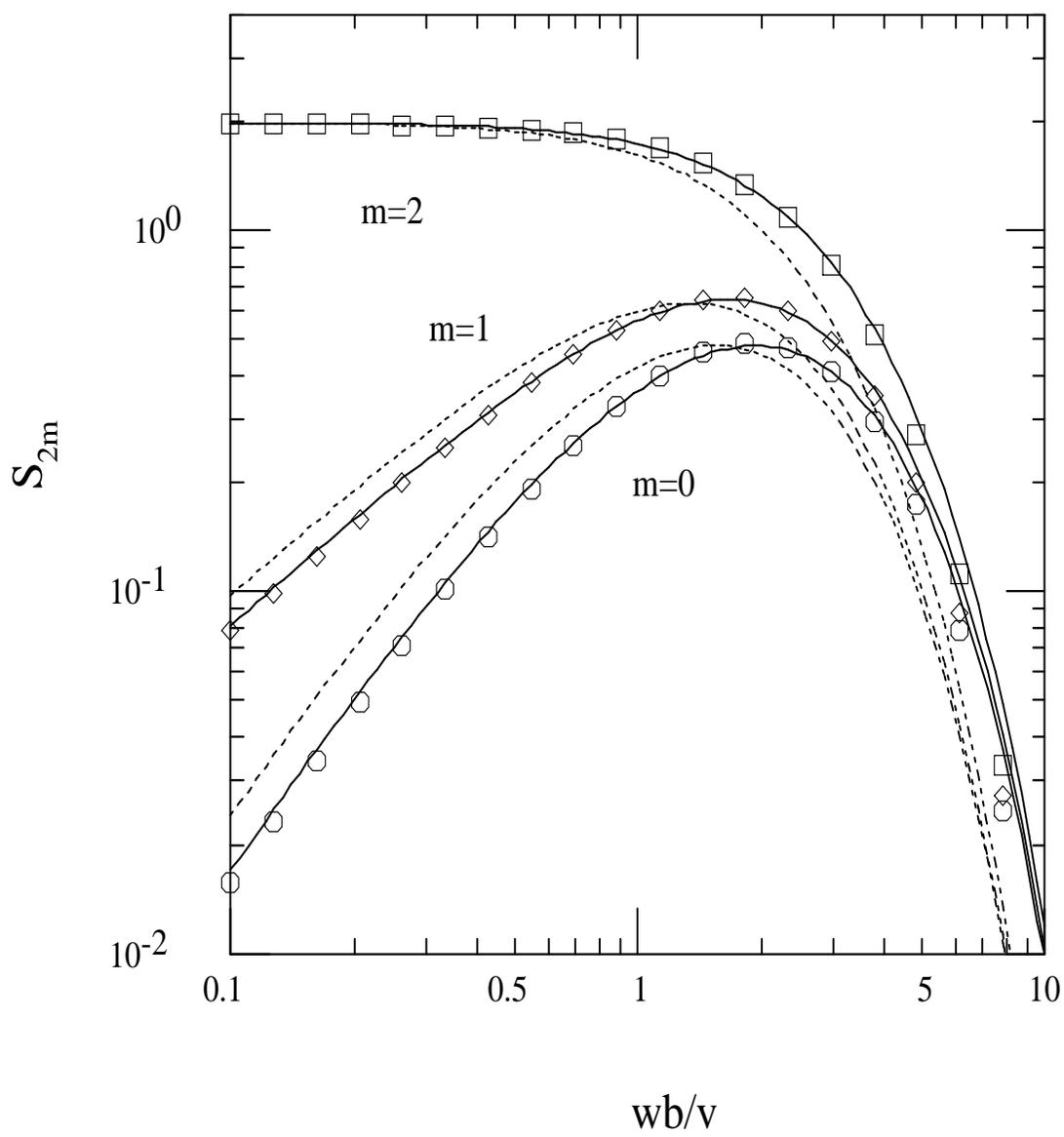
\begin{figure}
\caption{Similar to Fig. 2 but for quadrupole excitations.}
\label{autonum3}
\end{figure} 

\end{document}